\documentclass[sigconf]{acmart}

\usepackage{booktabs}
\usepackage{enumitem}
\usepackage{dblfloatfix}
\usepackage{multirow}
\usepackage{footnote}
\usepackage{array}
\usepackage{amsmath}
\usepackage{listings}
\usepackage{color}
\usepackage{colortbl}
\usepackage{balance}
\usepackage{rotating}
\usepackage{tablefootnote}
\usepackage{ifthen}
\usepackage{tikz}

\setitemize{leftmargin=*}

\newif\ifIsDoubleBlind
\IsDoubleBlindfalse

\definecolor{ColorWeakLightBlue}{RGB}{220, 240, 255}
\definecolor{ColorHighlightLightBlue}{RGB}{191, 225, 255}
\definecolor{ColorLightGray}{RGB}{220, 220, 220}

\newcolumntype{L}[1]{>{\raggedright\let\newline\\\arraybackslash\hspace{0pt}}m{#1}}
\newcolumntype{R}[1]{>{\raggedleft\let\newline\\\arraybackslash\hspace{0pt}}m{#1}}

\newcolumntype{Z}[1]{>{\raggedright\let\newline\\\arraybackslash\hspace{0pt}}b{#1}}

\newcommand{\formatMetricIndex}[1]{M#1}
\newcommand{\formatItemName}[1]{\mbox{\textit{#1}}}
\newcommand{\formatCaptionDetails}[1]{\textmd{\small{#1}}}

\newcommand{\coloredCircle}[1]{\tikz\draw[#1,fill=#1] (0,0) circle (.5ex);}

\begin{document}

\copyrightyear{2018}
\acmYear{2018}
\setcopyright{rightsretained}
\acmConference[ICSE '18 Companion]{40th International Conference on Software Engineering Companion}{May 27-June 3, 2018}{Gothenburg, Sweden}
\acmBooktitle{ICSE '18 Companion: 40th International Conference on Software Engineering Companion, May 27-June 3, 2018, Gothenburg, Sweden}
\acmDOI{10.1145/3183440.3195022}
\acmISBN{978-1-4503-5663-3/18/05}

\title{Poster: Identification of Methods with Low Fault Risk}

\ifIsDoubleBlind
\else
	\author{Rainer Niedermayr}
\affiliation{
  \institution{University of Stuttgart, CQSE GmbH}
  \city{Garching b. M\"unchen} 
  \state{Germany} 
}
\email{niedermayr@cqse.eu}

\author{Tobias R\"ohm}
\affiliation{
  \institution{CQSE GmbH}
  \city{Garching b. M\"unchen} 
  \country{Germany}}
\email{roehm@cqse.eu}

\author{Stefan Wagner}
\affiliation{
  \institution{University of Stuttgart}
  \city{Stuttgart} 
  \state{Germany} 
}
\email{stefan.wagner@informatik.uni-stuttgart.de}

\renewcommand{\shortauthors}{R. Niedermayr et al.}
\fi

\ifIsDoubleBlind
\else
\fi

\newcommand{\topNRules}{165}
\newcommand{\topNRulesLowPrec}{336}
\newcommand{\minSuppVal}{10\%}
\newcommand{\minConfVal}{90\%}

\begin{abstract}
Test resources are usually limited and therefore it is often not possible to completely test an application before a release.
Therefore, testers need to focus their activities on the relevant code regions.
In this paper, we introduce an inverse defect prediction approach to identify methods that contain hardly any faults.
We applied our approach to six Java open-source projects
 and show that on average 31.6\% of the methods of a project have a low fault risk;
 they contain in total, on average, only 5.8\% of all faults.
Furthermore, the results suggest that, unlike defect prediction, our approach can also be applied in cross-project prediction scenarios.
Therefore, inverse defect prediction can help prioritize untested code areas and guide testers to increase the fault detection probability.
\end{abstract}

\maketitle

\section{Introduction}
\label{Sec:Introduction}

Software testing can be very time consuming and test resources are usually scarce.
Therefore, it is often not possible to completely test the whole code base before each release.
Consequently, development teams must limit their testing scope and focus on code regions that have the best cost-benefit ratio regarding test resources~\cite{hall2012systematic}.

To support development teams in this activity, defect prediction has been developed and studied extensively.
Defect prediction identifies code regions that are likely to contain a fault and should therefore be tested~\cite{menzies2007data, Weyuker2008FaultPrediction}.
However, after several decades of research on defect prediction, it is still hardly used in practice.
Defect prediction models need to be trained with precise historical fault data from the project, which is often not available---especially in new projects.
Cross-project predictions, which use models trained from data of other projects, are still considered as a difficult task~\cite{zimmermann2009cross, turhan2009relative}.

This paper suggests an alternative approach for prioritizing code regions: inverse defect prediction (IDP).
The idea behind IDP is to identify code regions with \textit{low} fault risk,
 which can be deferred when writing automated tests if none yet exist.
The main difference to traditional defect prediction lies in the predicted classes
 and in the optimization target.
While defect prediction classifies an artifact either as buggy or non-buggy,
 IDP identifies artifacts that exhibit a low fault risk
 and does not make an assumption about the remaining artifacts.
Instead of predicting all non-faulty artifacts,
 IDP aims to identify only those that contain \textit{no faults with high certainty.}
Therefore, IDP strives to achieve a high precision for the identified artifacts,
 while recall is less important.
In contrast, defect prediction aims at a high recall to detect as many faults as possible
 and at a high precision such that only few false positives occur.

We applied IDP on the Defects4J dataset~\cite{just2014defects4j} at the method level.
We evaluated how many faults the identified low-fault-risk methods contain
 and how much savings potential can be gained by ignoring them during testing.
Our results show that IDP can successfully identify low-fault-risk methods,
 which contain considerably less faults than an arbitrary method,
 and are frequent enough to provide a worthwhile savings potential for QA activities.
Moreover, our results indicate that IDP can be an alternative to traditional defect prediction in cross-project prediction scenarios.
\section{Approach}
\label{Sec:Approach}
The IDP approach to identify low-fault-risk methods comprises the computation of source-code metrics for each method,
 the data pre-processing before the mining, and the creation of a classifier using association rule mining.

Like defect prediction models, IDP needs metrics to train a classifier
 and an indication whether a method was faulty at least once.
We computed for each method common software-analysis metrics,
 metrics that count occurrences of Java language constructs,
 and categories to which a method can belong to.
Table~\ref{Tbl:Metric_Set_Short} presents an excerpt of the 39 used metrics.


\begin{table}
	\small
	\centering
	\caption{Computed metrics for each method (excerpt).}
	\begin{tabular}{rll}
		\toprule
														 & Metric Name & Type \\
		\midrule
			 \formatMetricIndex{1} & Source Lines of Code (SLOC) & length \\
			 \formatMetricIndex{2} & Cyclomatic Complexity (CC) & complexity \\
			 \formatMetricIndex{3} & Max. Nesting Depth & max. value \\
			 \formatMetricIndex{9} & Array Accesses & count \\
			\formatMetricIndex{11} & Value Assignments & count \\
			\formatMetricIndex{16} & If Conditions & count \\
			\formatMetricIndex{20} & Loops & count \\
			\formatMetricIndex{21} & Method Invocations & count \\
			\formatMetricIndex{23} & Null Literals & count \\
			\formatMetricIndex{24} & Return Statements & count \\
			\formatMetricIndex{29} & Ternary Operations & count \\
			\formatMetricIndex{30} & Throw Statements & count \\
			\formatMetricIndex{34} & Is Constructor & category \\
			\formatMetricIndex{36} & Is Getter & category \\
			\tiny{\dots} & & \\
		\bottomrule
	\end{tabular}
	\label{Tbl:Metric_Set_Short}
\end{table}

As association rule mining computes frequent itemsets from categorical attributes,
 our next step was to discretize the numerical metrics.
For the numeric software-analysis metrics, such as SLOC and the cyclomatic complexity,
 we inspected the value distribution, computed tertiles,
 and created a new binary metric for each tertile.
For all count metrics, we created a binary ``has-no''-metric,
  which is true if the value is zero, e.g.,
 \mbox{\formatItemName{CountLoops} $= 0$} \mbox{$\Rightarrow\ $ \formatItemName{NoLoops} $=$ \texttt{true}}.
For the method categories (setter, getter, \dots), no transformation was necessary since they are already binary.

Defect datasets are often highly imbalanced and contain only a small portion of faulty methods~\cite{khoshgoftaar2010attribute}.
Therefore, we applied SMOTE\footnote{Synthetic Minority Over-sampling Technique},
 a well-known algorithm for over- and under-sampling,
 to the training data to address the imbalance.

To identify low-fault-risk methods, we first mined association rules of the type
 $\{$\formatItemName{Metric1}, \formatItemName{Metric2}, \formatItemName{Metric3}, \dots$\} \rightarrow \{$\formatItemName{NotFaulty}$\}$
 by applying the \textit{Apriori} algorithm~\cite{agrawal1994fast} to the training data.
Next, we ordered the obtained rules descending by their confidence value.
The \textit{confidence} of a rule expresses
 the proportion of methods that are non-faulty and satisfy the metric predicates out of all methods that satisfy the metric predicates.
Thus, it can be considered as the precision of a rule.
Finally, to build the low-fault-risk classifier,
 we combined the top \textit{n} association rules with the highest confidence values using the logical-or operator.
Hence, we considered a method to have a low fault risk if at least one of the top \textit{n} rules matched.
To determine \textit{n}, we computed the maximum number of rules
 until the share of faulty methods in low-fault-risk methods of the training set exceeded a certain threshold.

\section{Evaluation}
\label{Sec:Empirical_Study}

We evaluated our approach on the Defects4J dataset~\cite{just2014defects4j},
 which contains real faults for six open-source projects.
We performed both within- and cross-project predictions and evaluated the identified methods.
For within-project predictions,
 we applied 10-fold cross-validation.
We randomly sampled the dataset of each project into ten stratified partitions of equal size
 and used each sample once for testing the classifier, which is trained on the remaining nine samples.
Table~\ref{Tbl:TopRulesShort} exemplarily presents resulting association rules.
For cross-project prediction,
 we evaluated each of the six projects with a classifier trained on data from the respective other five projects.

The results of the within-project predictions show that
 IDP classified between 16\% and 75.3\% of the methods as ``low fault risk'' (median: 31.6\%);
 the identified methods comprise between 5\% and 68.5\% of the SLOC (median: 14.2\%).
These methods are on average six times less likely to contain a fault than an arbitrary method.
Low-fault-risk methods are usually short,
 but their source code lines of code (SLOC) are still 3.4 times less likely to contain a fault.
Figure~\ref{Fig:ComparisonWithinAndCrossProject} illustrates the within-project results.

Cross-project IDP found between 18.4\% and 31.2\% of the methods to have a low fault risk (median: 23.2\%);
 they contain between 6.3\% and 10.1\% of the SLOC (median: 7.8\%).
Surprisingly, their fault density is lower (compared to within-project predictions),
 because they are nearly 11 times less likely to contain a fault than an arbitrary method.
Based on SLOC, they are 3.7 times less likely to contain a fault.


\begin{table}
	\centering
	\caption{Top three association rules for the project \textit{Apache Commons Lang} \formatCaptionDetails{(within-project prediction, fold 1).}}
	\begin{tabular}{Z{0.5cm}L{7.0cm}}
		\toprule
			\# & Rule \\
		\midrule

		1 & \small{
		 \{\formatItemName{UniqueVariableIdentifiersLessThan2},
     \formatItemName{NoMethodInvocations}\}
		 $\Rightarrow$~\{\formatItemName{NotFaulty}\}} \\

		2 & \small{
		 \{\formatItemName{SlocLessThan4},
     \formatItemName{NoAnyArithmeticOp},
     \formatItemName{NoMethodInvocations}\}
		 $\Rightarrow$~\{\formatItemName{NotFaulty}\}} \\

		3 & \small{
		 \{\formatItemName{SlocLessThan4},
     \formatItemName{MaxMethodChainingLessThan2},
     \formatItemName{UniqueVariableIdentifiersLessThan2}\}
		 $\Rightarrow$~\{\formatItemName{NotFaulty}\}} \\
				
		\bottomrule
	\end{tabular}
	\label{Tbl:TopRulesShort}
\end{table}
\definecolor{Color_IDP_Extent}{RGB}{0, 99, 71}
\definecolor{Color_IDP_FaultShare}{RGB}{255, 99, 71}

\begin{figure}
	\centering
	\includegraphics[width=1.0\linewidth,clip]{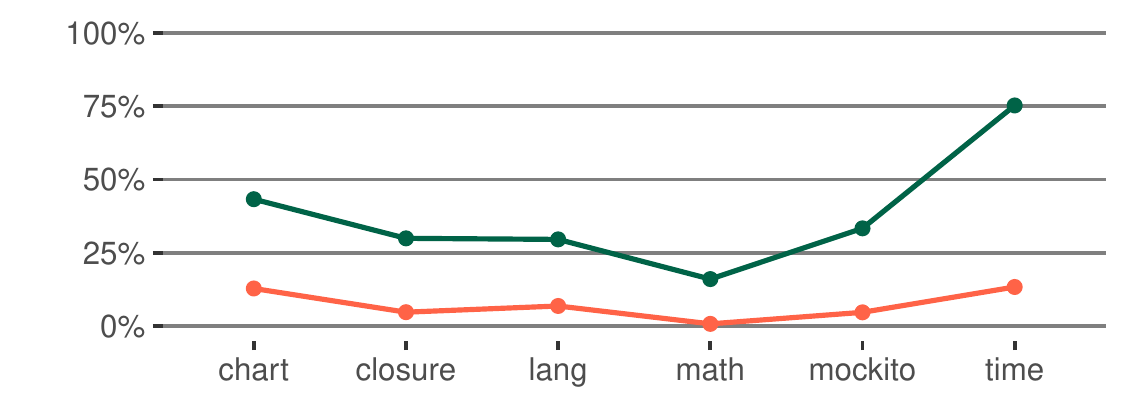}
	\caption[Proportion of low-fault-risk methods and their share in faults of all faults.]{
		\coloredCircle{Color_IDP_Extent}~Proportion of low-fault-risk methods classified by within-project IDP
		 and \coloredCircle{Color_IDP_FaultShare}~their share in faults of all faults.
		\formatCaptionDetails{
		 In the project \textit{Google Closure Compiler},
		  IDP classified 29.9\% of the methods to have a low fault risk;
			they contain only 4.7\% of all faults.
		}
	}
	\label{Fig:ComparisonWithinAndCrossProject}
\end{figure}
\section{Conclusion}
\label{Sec:Conclusion}
Inverse defect prediction using association rule mining can be used to identify low-fault-risk methods.
The identified methods are indeed considerably less likely to contain a fault
 and can provide a worthwhile savings potential for QA activities.
This applies to both within- and cross-project predictions.

\ifIsDoubleBlind
\else
	\begin{acks}
	\small{
		This work was partially funded by the German Federal Ministry of Education and Research (BMBF), grant ``Q-Effekt, 01IS15003A''.
		The responsibility for this article lies with the authors.
	}
\end{acks}
\fi


\bibliographystyle{ACM-Reference-Format}
\bibliography{./literature/bibtex,./literature/bibtex_unused}

\end{document}